\documentclass[a4paper,12pt]{article}
\usepackage{epsfig}
\usepackage{rotating}
\usepackage{amsmath,amssymb}
\usepackage{graphicx}
\usepackage{setspace}
\usepackage{fancyhdr}
\usepackage{moreverb}
\usepackage{rotating}
\usepackage{cite}
\def\rp{$R_p \hspace{-1em}/\;\:$ }
\def\ti{\tilde}
\def\ed{\end{document}}

\def\stau{\tilde \tau}
\def\smu{\tilde \mu}
\def\se{\tilde e}

\newcommand{\lsim}{\;\raisebox{-0.9ex}{$\textstyle\stackrel{\textstyle<}
          {\sim}$}\;}

\def\21{$SU(2) \otimes U(1) $}

\newcommand{\AddrVienna}{
\it  Institut f\"ur Theoretische Physik, Universit\"at Wien, \\ 
A-1090 Vienna, Austria \\}

\newcommand{\AddrAHEP}{%
 \it AHEP Group, Instituto de F\'{\i}sica Corpuscular --
  C.S.I.C.\\ Universitat de Val{\`e}ncia \\
  Edificio de Institutos de Paterna, Apartado 22085,
  E--46071 Val{\`e}ncia, Spain\\}

\newcommand{\AddrETH}{%
 \it Institut f\"ur Theoretische Physik, Universit\"at Z\"urich, \\ 
CH-8057 Z\"urich, Switzerland}

\begin{document}

\begin{flushright}hep-ph/0306071\\  IFIC/03-24\\  ZU-TH 07/03 \\
    UWThPh-2003-05
  \end{flushright}  

\begin{center}  
  \textbf{\large 
Testing the Mechanism of R-parity Breaking with Slepton LSP Decays}\\[10mm]

{A. Bartl${}^1$, M. Hirsch${}^2$, T. Kernreiter$^{1,2}$, W. Porod${}^3$ and 
J. W. F. Valle${}^2$ } 
\vspace{0.3cm}\\ 

$^1$ \AddrVienna
$^2$ \AddrAHEP
$^3$  \AddrETH
\end{center}



\begin{abstract}
  In supersymmetric models R-parity can be violated through either
  bilinear or trilinear terms in the superpotential, or both. If
  charged scalar leptons are the lightest supersymmetric particles,
  their decay properties can be used to obtain information about the
  relative importance of these couplings. We show that in some
  specific scenarios it is even possible to decide whether bilinear or
  trilinear terms give the dominant contribution to the neutrino mass
  matrix.

\end{abstract}

\newpage

\section{Introduction}

The most general form for the R-parity violating and lepton number 
violating part of the superpotential is given by

\begin{equation}
W_{R_p \hspace{-0.8em}/\;\:}=\varepsilon_{ab}\left[ 
\frac{1}{2}\lambda_{ijk}\widehat L_i^a\widehat L_j^b\widehat E_k 
+\lambda'_{ijk}\widehat L_i^a\widehat Q_j^b\widehat D_k 
+\epsilon_i\widehat L_i^a\widehat H_u^b\right]~.
\label{eq:rpvpot}
\end{equation}
Taking into account the asymmetry of $\lambda_{ijk}$ in $i$ and $j$, 
Eq. (\ref{eq:rpvpot}) contains nine different $\lambda_{ijk}$, 
27 $\lambda'_{ijk}$ and three $\epsilon_i$, for a total of 39 parameters. 
In principle - as already pointed out in \cite{Hall:1983id} - the number 
of parameters in Eq. (\ref{eq:rpvpot}) can be reduced by 3 by a suitable 
rotation of basis and a number of authors have used this freedom 
to absorb the bilinear parameters $\epsilon_i$ into re-defined 
trilinear parameters \cite{dreiner:1997uz}.

However, for consistency one has to add three more bilinear terms 
to $V_{soft}$, the SUSY breaking potential, 
\footnote{We will not consider the corresponding lepton number violating 
trilinear soft masses, $A_{ijk}$ and $A'_{ijk}$, since they will not 
affect our conclusions.} 
\begin{equation}
V_{\rm{soft},R_p \hspace{-0.8em}/\;\:} = -\varepsilon_{ab}
          B_i\epsilon_i\widetilde L_i^aH_u^b \, .
\label{eq:rpvsoft}
\end{equation}
It is important to note that there is no rotation which can eliminate 
the bilinear terms in Eq. (\ref{eq:rpvsoft}) and Eq. (\ref{eq:rpvpot}) 
simultaneously \cite{Hall:1983id}  
\footnote{Unless the parameters $B_i$ and the corrsponding MSSM parameter 
$B$ are equal (as well as $m_{L_i}^2 \equiv m_{H_d}^2$). However, such an 
equality is unstable under RGE running \cite{decarlos:1997yh}.} and 
thus we are back to 39 independent parameters.

With this huge number of parameters it seems hardly possible to make any 
definite prediction for phenomenology. Could one reduce the number of free 
parameters in the theory? On theoretical grounds one could argue that a 
model with only bilinear terms is a self-consistent theory, whereas a 
model with only trilinear terms is not. The reasoning behind this argument 
is quite simple: Renormalization group running \cite{decarlos:1997yh}
will generate bilinears whenever trilinears are present, but RGEs which 
start with bilinears only will never generate trilinears~\cite{Diaz:1999is}. 
In addition, models which break R-parity spontaneously \cite{SBRPVold}
at the TeV scale \cite{SBRPV} will lead to effective low-energy
models where only bilinear terms are present.

On the other hand, experiments are done in the physical mass-eigenstate 
basis. Before exploring the implications of Eq. (\ref{eq:rpvpot}), one 
therefore has to re-diagonalize all mass matrices of the model.  The 
essential point here is that the bilinears introduce mixing among the 
various states of the model and thus lead to the appearance of
``effective'' trilinears in the mass basis, for example,
\begin{equation} \label{lpeff}
\lambda'_{333} \sim \frac{\epsilon_3}{\mu}h_b
\end{equation}
where $h_b$ and $\mu$ are the bottom Yukawa coupling and the Higgs
mixing parameter. These ``effective'' trilinears are unavoidable even 
in models which start with bilinear terms in the superpotential only. 
The essential point to realize here, however, is that these  ``effective'' 
trilinears necessarily always follow the hierarchy implied by the 
Yukawa couplings of the standard model and therefore are {\em not} new 
free parameters of the theory in {\em contrast} to ``genuine'' trilinears. 

All terms in Eq. (\ref{eq:rpvpot}) violate lepton number by one unit and 
thus will necessarily contribute to the (Majorana) neutrino mass matrix. 
Quite a number of articles have investigated the consequences of 
Eq. (\ref{eq:rpvpot}) for neutrino physics, some of them considering only 
the bilinear terms \cite{hirsch:2000ef},\cite{hempfling:1996wj} others 
only the trilinear terms \cite{drees:1998id} and a few have also entertained 
the possibility that both kind of terms exist \cite{davidson:2000ne}.

An obvious question to ask then is: Which of the terms in 
Eq. (\ref{eq:rpvpot}) give the dominant contribution to the neutrino 
masses? And, is there {\em any} possibility to settle this question 
{\em experimentally}?
The aim of this paper is to demonstrate that, if charged scalar 
leptons are the lightest supersymmetric particles (LSPs), there are 
different broken R parity scenarios where one can probe for the origin 
of neutrino mass.

How likely is it that charged sleptons are the LSPs? In models of
supersymmetry breaking based on minimal supergravity boundary
conditions (mSugra) \cite{Haber:1985rc}, in gauge mediated SUSY breaking
\cite{Giudice:1998bp} and in anomaly mediated SUSY breaking
\cite{Giudice:1998xp,Randall:1998uk} this actually happens in sizeable
portions of the parameter space. Usually this possibility is declared 
``ruled out cosmologically'' and simply discarded (except in GMSB,
where the charged sleptons are actually the NLSPs). However, with
R-parity violated the charged scalars decay, so cosmological limits
simply do not apply.~\footnote{A possible dark matter candidate in this
  case is the axion.}

It is exactly the observation that the LSP decays through the same
R-parity violating operators, which also govern the entries in the
neutrino mass matrix, which forms the basic idea of the current paper.
To give a trivial example, in a world were only $\lambda_{ijk}$'s are
non-zero, charged scalars can decay only to leptonic final states,
whereas if only $\lambda'_{ijk}$'s are non-zero, there will only be
hadronic final states. Of course, this example is vastly oversimplied,
but here we will argue that, with experimental information from
neutrino physics \cite{Eguchi:2002dm}, it is possible to fix the size of 
the various couplings assuming one or the other combination being dominant 
and check the consequences for the charged slepton decays.

Upper limits on all parameters in Eq. (\ref{eq:rpvpot}) exist and 
have been extensively discussed in the literature 
\cite{dreiner:1997uz,barbier:1998fe}, \cite{dreiner:2001kc,hirsch:1996ek}.  
However, in agreement with \cite{Rakshit:1998kd}, we find that limits on 
the couplings imposed by current neutrino oscillation data are usually 
much stronger than all other indirect limits.  
\footnote{This claim is true only if
  the left-right mixing in the squark/slepton sector is not exactly
  zero.  Exactly vanishing left-right mixing can not be excluded at
  present, but requires fine tuning of parameters. We will not
  consider this possibility.}  
Notable exceptions from this general rule are the limits from double 
beta decay \cite{hirsch:1996ek}, from non-observation of $\mu \to  3 e$, 
$\mu Ti \to  e Ti$, and from $\Delta m_B$ and $\Delta m_K$
\cite{barbier:1998fe,Rakshit:1998kd}.
Here we note only that none of the existing limits is strong enough to
require charged scalar leptons decaying with visible decay lengths. 

Scalar tau LSPs have been discussed in the literature before. 
In \cite{Akeroyd:1997iq} scalar tau decays within bilinear R-parity 
violation have been discussed for a one generation model of R-parity 
violation. In \cite{Akeroyd:2001pm} possible similarities between 
the phenomenology of a charged Higgs and a stau have been discussed 
within trilinear R-parity violation. In \cite{Hirsch:2002ys} charged 
scalar LSP decays within bilinear R-parity violation (with 3 generations) 
have been discussed with special emphasis on their relation to neutrino 
physics. 
\footnote{After completion of this article, the preprint 
\cite{Allanach:2003eb} appeared. The authors study general R-parity 
violation in mSugra considering also the case of a scalar tau LSP. 
However, the emphasis of  \cite{Allanach:2003eb} is on the supersymmetry 
breaking parameters and no attempt is made to explain current neutrino 
data by the R-parity violating parameters.}
And in GMSB the scalar tau can be a quite long-lived 
next-to-lightest supersymmetric particle (NLSP), which from the collider 
point of view looks like a stable, charged LSP, see for 
example \cite{Giudice:1998bp}. Finally charged scalar LSPs have been 
discussed in an R-parity {\em conserving} extension of the MSSM in 
\cite{deGouvea:1998yp}. These authors  argue 
that one should not take into account the cosmological limits on 
charged LSPs, because alternative cosmologies can provide loopholes 
to the usual ``overclosure'' argument \cite{deGouvea:1998yp}. 

This paper is organized as follows: In the next section we will set up
the model definitions. In section 3 the charged scalar decay widths
are given. Section 4 summarizes some necessary formulas for the
neutrino mass matrix, while section 5 discusses charged scalar lepton
decays under different hypotheses both analytically and numerically.
We then close with a short summary and outlook.

\section{The Model}

We work in the minimally supersymmetrized version of the
standard model (SM) \cite{Haber:1985rc}, augmented with the R-parity 
violating terms in Eq. (\ref{eq:rpvpot}). Superpotential and 
mass matrices of the MSSM have been given exhaustively in the 
literature and will not be repeated here. For mass matrices 
including the bilinear R-parity violating terms see, for example 
\cite{hirsch:2000ef}.

As will be seen in the next section, the slepton decay widths 
depend on left-right mixing in the scalar sector. We therefore give 
a short repetition of $\ti\ell_L$ -- $\ti\ell_R$ mixing.
The masses and couplings of the
$\ti\ell$ follow from the 
symmetric $2 \times 2$ mass matrix which in the basis 
$(\ti\ell_R, \ti\ell_L)$ reads \cite{Haber:1985rc,ellis}
\begin{equation} \label{eq:mm}
{\mathcal{L}}_M^{\ti\ell}= -(\ti\ell_R^{\dagger},\, \ti\ell_L^{\dagger})
\left(\begin{array}{ccc}
M_{\ti\ell_{RR}}^2 & M_{\ti\ell_{RL}}^2 \\[2mm]
M_{\ti\ell_{LR}}^2 & M_{\ti\ell_{LL}}^2
\end{array}\right)\left(
\begin{array}{ccc}
\ti\ell_R\\[2mm]
\ti\ell_L \end{array}\right),
\end{equation}
where
\begin{eqnarray}
M_{\ti\ell_{LL}}^2 & = & M_{\tilde L}^2+
m_Z^2 \cos2\beta \left(\sin^2\Theta_W-\frac{1}{2}\right)
+m_{\ell}^2 ,\\[3mm]
M_{\ti\ell_{RR}}^2 & = & M_{\tilde E}^2-
m_Z^2\sin^2\Theta_W\cos2\beta + m_{\ell}^2 ,\\[3mm]
M_{\ti\ell_{RL}}^2 & = & M_{\ti\ell_{LR}}^2=
m_{\ell}(A_{\ell}-\mu \tan\beta).
\label{eq:mlr}
\end{eqnarray} 
Here $\tan\beta=v_u/v_d$ with $v_d (v_u)$ being the vacuum 
expectation value of the Higgs field $H_d^0 (H_u^0)$,
$m_{\ell}$ is the mass of the appropriate lepton and 
$\Theta_W$ is the weak mixing angle, $\mu$ is the 
Higgs--Higgsino mass parameter and $M_{\ti L}$, 
$M_{\ti E}, A_{\ell}$ are the soft SUSY--breaking parameters of 
the slepton system\cite{Haber:1985rc}.  
The mass eigenstates $\ti\ell_i$ are $(\ti\ell_1, \ti\ell_2)=
(\ti\ell_R, \ti\ell_L) {{\mathcal R}^{\ti\ell}}^T$ with
\begin{equation} \label{eq:rtau}
{\mathcal R}^{\ti\ell}=\left( \begin{array}{ccc}
\cos\theta_{\ti\ell} & \sin\theta_{\ti\ell}\\[2mm]
-\sin\theta_{\ti\ell} & \cos\theta_{\ti\ell}
\end{array}\right),
\end{equation}
with
\begin{equation} \label{thtau}
\cos\theta_{\ti\ell}=-\frac{M_{\ti\ell_{LR}}^2}
{\sqrt{(M_{\ti\ell _{LR}}^2)^2+
(m_{\ti\ell_1}^2-M_{\ti\ell_{RR}}^2)^2}}~,\
\sin\theta_{\ti\ell}=\frac{M_{\ti\ell_{RR}}^2-m_{\ti\ell_1}^2}
{\sqrt{(M_{\ti\ell_{LR}}^2)^2+(m_{\ti\ell_1}^2-M_{\ti\ell_{RR}}^2)^2}}
\end{equation}
\\
while the mass eigenvalues are given by
\begin{equation} \label{eq:m12}
m_{\ti\ell_{1,2}}^2 = \frac{1}{2}\left((M_{\ti\ell_{LL}}^2+
M_{\ti\ell_{RR}}^2)\mp 
\sqrt{(M_{\ti\ell_{LL}}^2 - M_{\ti\ell_{RR}}^2)^2 +
4 (M_{\ti\ell_{LR}}^2)^2}\right).
\end{equation}
Due to the appearance of the lepton mass in Eq. (\ref{eq:mlr}) one 
expects the left-right mixing is tiny in the seletron, small in 
the smuon and potentially sizeable in the stau sectors. 

In the R-parity violating version of the MSSM a priori any SUSY 
particle can be the LSP. However, supplementing the MSSM with mSugra 
motivated boundary conditions, one usually finds that either the 
lightest neutralino or one of the charged scalars is the LSP (CSLSP). 
As a rule of thumb, charged scalars are LSPs if $m_0 \ll M_{1/2}$ and 
$\mu$ large. CSLSPs in these models are usually mainly right sleptons. 
Furthermore, even though one expects some splitting between ${\tilde \tau}$, 
${\tilde \mu}$ and ${\tilde e}$ from RGE running, the latter are 
not much heavier than the former and so might also dominantly 
decay through R-parity violating operators even though they are 
not strictly the LSP. \footnote{This is similar to what happens 
quite generically in GMSB models, where such a scenario is called 
co-NLSP \cite{Giudice:1998bp}.}

\section{Decay Widths}

Charged scalar leptons lighter than all other supersymmetric particles 
will decay through \rp vertices. Possible final states are either 
$\ell_i\nu_j$ or $q\bar q'$. We will present the decay widths 
taking into account only the trilinear parameters. For the corresponding 
formulas for the bilinear terms see \cite{Hirsch:2002ys}.

The relevant Lagrangian to study these two-body decays of $\ti\ell_1$ is 
obtained from Eq.~(\ref{eq:rpvpot}). It is given by \cite{dreiner:1997uz},
\begin{eqnarray}\label{eq:lagrange}
{\mathcal L}_{\rm int}&=& 
\ti e_{j1}{\bar e}_{k}\lbrack 
(\sin\theta_{\tilde e_j}\lambda_{ijk}) P_L+
(\cos\theta_{\tilde e_j}\lambda_{ikj}) P_R \rbrack \nu_{i}
\nonumber\\[2mm]
&&{}
-\sum_j V_{nj}\lambda'_{ijk}\sin\theta_{\tilde e_i}
\ti e_{i1}\bar d_k P_L u_n+{\rm h.c.}~,
\end{eqnarray}
where $P_{R,L}=(1\pm\gamma_5)/2$ and $V_{nj}$ is the corresponding 
element of the CKM matrix. Here we work in a basis where the CKM 
matrix is solely due to the mixing of the up-type quarks. 
[$\ti e_{j1}$ denotes the lighter slepton mass eigenstate with
family index $j$].
In the limit $m_{\ell_j}\ll m_{\ti\ell_1}$ 
the leptonic two--body decay widths of $\ti\ell_1$
read,
\begin{equation} \label{eq:lep}
\Gamma(\ti e_{j1} \to e_{k}\sum_i \nu_{i})=
\frac{m_{\ti e_{j1}}}{16\pi}\sum_i\lbrack
(\sin\theta_{\tilde e_j}\lambda_{ijk})^2+
(\cos\theta_{\tilde e_j}\lambda_{ikj})^2
\rbrack
\end{equation}
For $\ti\ell_1\simeq \ti\ell_R$, we find for the branching ratios
\begin{equation} \label{eq:brlep}
Br^{\ti\ell}_{(1,2,3)} \simeq 
\frac{1}{2}\left[1-\frac{(\lambda_{23\ti\ell}^2,\lambda_{13\ti\ell}^2,
\lambda_{12\ti\ell}^2)}{\sum_{i<j}\lambda_{ij\ti\ell}^2}
\right]~,
\end{equation}
where we have introduced the following shorthand notation:
$Br^{\ti\ell}_{(1,2,3)}\equiv Br(\ti\ell_1\to (e,\mu,\tau) \sum \nu_i)$.
Corrections to Eq.~\eqref{eq:brlep} are $\propto \theta_{\ti\ell}^2 \lll 1$. 
In the Appendix we give the solutions to Eq.~\eqref{eq:lep} with respect to 
the trilinear couplings $\lambda^2_{ijk}$ by making an expansion in the 
small parameters $\theta_{\ti\ell}$. 
In the limit $\theta_{\ti\ell} \to  0$ , we derive form 
Eq.~\eqref{eq:brlep} that
 
\begin{equation}
Br^{\ti\ell}_i < 0.5,~\forall~i
\label{OneHalf}
\end{equation}
It is interesting to note that Eq. (\ref{OneHalf}) is a {\em definite
  prediction} (correct up to order $\theta_{\ti\ell}^2$) {\em of
  trilinear R-parity violation} despite the large number of
parameters. An experimental result contradicting Eq. (\ref{OneHalf})
would be a clear sign for bilinear R-parity breaking, if the CSLSPs
are mainly right (isosinglet) sleptons.

In order to determine $\lambda^2_{ijk}$ a measurement of the total width 
of the sleptons is necessary. Information on the widths could be obtained 
if a finite decay length is observed, we will return 
to this point later. However, because the next to leading term in 
Eqs \eqref{eq:tricoupe}-\eqref{eq:tricoupt} is only of the order 
$\theta_{\ti\ell}^2$ ratios $\lambda^2_{ijk}/\lambda^2_{i'j'k}$ may 
be determined without this information from the branching ratios 
$Br(\ti e_k\to e_l \sum \nu_m)$ alone.

Note also that there are 9 different $\lambda_{ijk}$ and 9 observables, 
which implies that in principle it is possible to determine all 
$\lambda_{ijk}$ if there is {\em no} bilinear R-parity breaking. 
Conversely, adding also the bilinear terms in the superpotential 
and/or in the soft SUSY breaking potential a full reconstruction of all 
parameters from the leptonic decays of the CSLSPs is impossible in 
principle. Differentiating between bilinear and trilinear 
R-parity breaking therefore necessarily requires the construction 
of conditions such as Eq. (\ref{OneHalf}).

For the hadronic two--body decay widths of $\ti\ell_1$ one finds
\begin{equation} \label{eq:hadwidth}
(\Gamma^{\ti\ell}_h)_{nk} \equiv \Gamma(\ti\ell_1\to \bar{u}_n d_k)=
N_c~\beta \ \frac{m_{\ti\ell}}{16\pi} 
\ \sin^2\theta_{\ti\ell} \ |\sum_j V_{nj} \ 
{\lambda'}_{\ti\ell jk}|^2~,
\end{equation}
where $N_c=3$ is the number of colours. $\beta=1$ for $j\neq 3$ and 
$\beta= (1-(\frac{m_t}{m_{\ti\ell}})^2 )^2$ for $j=3$.
For the expected small mixing in the first generation of sleptons the 
hadronic width is highly suppressed compared to the leptonic width if 
$\lambda'\lsim \lambda$.

\section{Neutrino Mass Matrix}

The Majorana mass term for the neutrinos is of the form
\begin{equation} \label{eq:massneu}
{\mathcal L}_{\rm mass}=-\frac{1}{2}~
(\nu_{Li})^T\ C~
(M_{\nu})_{ii'}~\nu_{Li'}+{\rm h.c.}~
\end{equation}
where $\nu_{Li}$ denote the left--handed weak eigenstates and
$M_{\nu}$ is a complex symmetric $3\times 3$
matrix~\cite{schechter:1980gr}.

Contributions to the neutrino mass matrix are induced at tree-level by
the bilinear \rp terms. They can be expressed as
\cite{hirsch:1998kc,Joshipura:ib}
\begin{eqnarray} \label{eq:meff}
m_{\rm eff}=\frac{M_1 g^2+M_2 {g'}^2}
{4\ \det({\mathcal M}_{\chi^0})}
\left(\hskip -2mm \begin{array}{ccc}
\Lambda_e^2 
\hskip -1pt&\hskip -1pt
\Lambda_e \Lambda_\mu
\hskip -1pt&\hskip -1pt
\Lambda_e \Lambda_\tau \\
\Lambda_e \Lambda_\mu 
\hskip -1pt&\hskip -1pt
\Lambda_\mu^2
\hskip -1pt&\hskip -1pt
\Lambda_\mu \Lambda_\tau \\
\Lambda_e \Lambda_\tau 
\hskip -1pt&\hskip -1pt 
\Lambda_\mu \Lambda_\tau 
\hskip -1pt&\hskip -1pt
\Lambda_\tau^2
\end{array}\hskip -3mm \right),
\end{eqnarray}
where $\Lambda_i=\mu v_i+v_d \epsilon_i$. Due to the projective nature of 
the effective neutrino mass matrix $ m_{\rm eff}$, only one neutrino 
acquires mass at the tree level. Therefore $m_{\rm eff}$ is diagonalized 
with only two mixing angles which can be expressed in terms of 
$\Lambda_i$:
\begin{equation} \label{eq:choozangle}
\tan\theta_{13} = - \frac{\Lambda_e}
                   {(\Lambda_{\mu}^2+\Lambda_{\tau}^2)^{\frac{1}{2}}}~,
\end{equation}
\begin{equation} \label{eq:atmangle}
\tan\theta_{23} = -\frac{\Lambda_{\mu}}{\Lambda_{\tau}}~.
\end{equation}
One-loop contributions to the neutrino mass matrix from bilinear terms
have been discussed extensively, see for example
\cite{hirsch:2000ef} and will not
be repeated here. At 1-loop level there are also contributions due to
trilinear \rp couplings. The relevant $\lambda$-terms are given in
Eq.~\eqref{eq:lagrange}, for $\lambda'_{ijk}$ type couplings we need:
\begin{equation} \label{eq:TriLagrange}
{\mathcal L}_{\lambda'}\supset
\lambda'_{ijk} \ \lbrack  
\ti d_{jL} {\bar d}_{k}P_L\nu_{i}+
{\ti d}^{\dagger}_{kR}{\bar \nu^c}_{i}P_L d_{j}
\rbrack+
{\rm h.c.}~.
\end{equation} 
The full mass matrix from trilinear terms is then given by 
$m^{1-loop}=m^{\lambda}+ m^{\lambda'}$, where
\begin{equation} \label{eq:Triloopmass}
m_{ii'}^{\lambda (\lambda')}=-\frac{1(N_c)}
{32\pi^2}~\lambda^{(')}_{ijk}\lambda^{(')}_{i'kj}
\left[
m_k\sin2\theta_j \ln\left(\frac{m^2_{2j}}{m^2_{1j}}\right)+
m_j\sin2\theta_k \ln\left(\frac{m^2_{2k}}{m^2_{1k}}\right)
\right]~,
\end{equation}
where $m_k$ is the appropriate fermion mass, $\theta_j$ denotes the
appropriate sfermion mixing angle and $m^2_{(1,2)j}$ are the
corresponding sfermion masses. Note the manifest symmetry of this
matrix as well as the presence of logarithmic factors.

We stress that Eq. \eqref{eq:Triloopmass} is an approximation to the
full 1-loop calculation. Numerically we have found that Eq.
\eqref{eq:Triloopmass} is good only up to factors of $\sim 2$ in
extreme cases. In the numerical part of this work we therefore always
diagonalize the full 1-loop corrected ($7 \times 7$)
neutrino-neutralino mass matrix.

\section{Neutrino Physics versus Collider Physics}

In the following we consider some specific scenarios with bilinear and
trilinear R-parity breaking terms being present in different
combinations and discuss the resulting decay patterns of the charged
sleptons. We will calculate the decay lengths 
\begin{equation}\label{eq:len}
L  =  \frac{\hbar c}{\Gamma}\sqrt{\frac{s}{4 m^2}-1}
\end{equation}
assuming a centre-of-mass
energy $\sqrt{s} = 0.8$~TeV, as typically could be reached in a future
linear collider
\cite{aguilar-saavedra:2001rg,Bartl:2003fh}.

In order to reduce the number of parameters, the numerical
calculations were performed in a constrained version of the MSSM. We
have scanned the parameters in the following ranges: $M_{1/2}\in
[0,1.2]$ TeV, $|\mu| \in [0,2.5]$ TeV, $m_0\in [0,1.2]$ TeV, $A_0/m_0$
and $B_0/m_0\in [-3,3]$ and $\tan\beta\in [2.5,10]$. All randomly
generated points were subsequently tested for consistency with the
stationary conditions of the scalar potencial as well as for
phenomenological constraints from supersymmetric particle searches 
\cite{pdg}. To be conservative we require the charged scalars to 
be heavier than 100 GeV, although existing limits are somewhat 
weaker and depend on the flavour of the charged scalar \cite{pdg}. 
We then select points in which charged scalars are the LSPs. This
latter cut strongly prefers $m_0 \ll M_{1/2}$ and large values of
$\mu$. Note again that our charged scalars are mainly right sleptons, as
discussed above. Also the low values of $\tan\beta$ in our scan lead 
to masses for the three different generations of charged sleptons
which are rather similar such that all three charged sleptons decay
dominantly through R-parity violating decay modes.

The \rp parameters were chosen in order to fulfill the requirements of 
the various scenarios to be discussed and as input from neutrino 
physics we have used, unless noted otherwise, the following ranges, 
as currently determined from neutrino oscillation experiments 
\cite{Eguchi:2002dm}:
\begin{equation} \label{eq:thatm}
0.3 < \sin^2\theta_{Atm}< 0.7
\end{equation}
\begin{equation} \label{eq:massatm}
1.2 \times 10^{-3}~ {\rm eV^2} < \Delta m_{Atm}^2 < 
4.8\times 10^{-3}~{\rm eV^2}
\end{equation}
\begin{equation} \label{eq:thsol}
0.29 < \tan^2\theta_{\odot}< 0.86
\end{equation}
\begin{equation} \label{eq:masssol}
5.1 \times 10^{-5}~ {\rm eV^2} < \Delta m_{\odot}^2 < 
1.9\times 10^{-4}~{\rm eV^2}
\end{equation}
\begin{equation} \label{eq:chooz}
\sin^2\theta_{13}< 0.05
\end{equation}

Before starting the discussion, let us briefly summarize the extreme
case in which neutrino oscillation parameters are determined only by
bilinear parameters~\cite{Hirsch:2002ys}.  Full details of this
particular situation can be found in \cite{Hirsch:2002ys} and will not
be repeated here. We will simply mention the following features
which are the most important tools to distinguish the pure bilinear
case from more complicated scenarios:

\begin{itemize}
\item The hierarchy in Yukawa couplings implies that the slepton decay
  lengths must be very different, $L(\stau_1) \ll L(\smu_1) \ll
  L(\se_1)$, in particular $L(\stau_1)/L(\smu_1) \sim
  (\frac{m_{\mu}}{m_{\tau}})^2$
\item Ratios of branching ratios must obey the following condition: 
\begin{eqnarray} \label{ratbil}
\frac{Br({\tilde \tau}_1 \to  e \sum \nu_i)/
Br({\tilde \tau}_1 \to  \mu \sum \nu_i)  }
{Br({\tilde \mu}_1 \to  e \sum \nu_i)/
Br({\tilde \mu}_1 \to  \tau \sum \nu_i)} \simeq
\frac{Br({\tilde e}_1 \to  \tau \sum \nu_i)}
{Br({\tilde e}_1 \to  \mu \sum \nu_i)}
\end{eqnarray}
\item Hadronic final states have branching ratios which are much too
  small to be observable
\item The $\se_1$ decays dominantly to the final state $e\sum\nu$ 
  (electrons plus missing (tranvserse) momentum)
\end{itemize}

Next we will discuss two scenarios with non-zero trilinears.

\subsection{Scenario I: $\lambda'_{ijk} \ll \lambda_{ijk}$}

Assume the atmospheric mass scale is generated by $m_{\rm eff}$ in
Eq.~(\ref{eq:meff}), whereas the solar mass scale is due to
$m^{\lambda}$ in Eq.~(\ref{eq:Triloopmass}).  If the couplings
$\lambda_{ij1}$ and $\lambda_{ij2}$ are not much larger than
$\lambda_{ij3}$ \footnote{Whether this is indeed the case may be
  checked, in principle, by a comparison of the decay length of the
  sleptons.}  the leading contribution to $m_{ii'}^{\lambda}$ comes
from the $\ti\tau_i-\tau$ loop and the contributions of all other
$\ti\ell_i-\ell$ loops are subleading. Then $m_{ii'}^{\lambda}$ is
given approximately as:
\begin{eqnarray} \label{eq:mlambda}
m^{\lambda}\approx -\frac{1}
{16\pi^2} m_{\tau}
\sin 2\theta_{\ti\tau}
\ln\left(\frac{m^2_{{\ti\tau}_2}}{m^2_{{\ti\tau}_1}}\right)
\left(\hskip -2mm \begin{array}{ccc}
\lambda^2_{133} 
\hskip -1pt&\hskip -1pt
\lambda_{133} \lambda_{233}
\hskip -1pt&\hskip -1pt
0~ \\
\lambda_{133}\lambda_{233}
\hskip -1pt&\hskip -1pt
\lambda^2_{233} 
\hskip -1pt&\hskip -1pt
0~ \\
0 
\hskip -1pt&\hskip -1pt 
0
\hskip -1pt&\hskip -1pt
0~
\end{array}\hskip -3mm \right)~.
\end{eqnarray}

\begin{figure}[t]
    \includegraphics[width=0.9\textwidth]{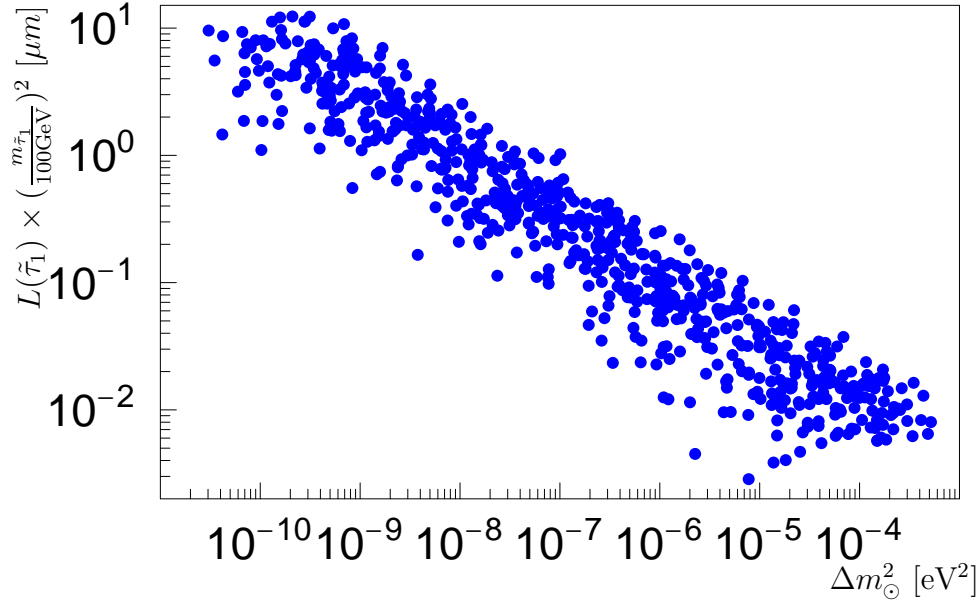}
\vskip-40mm
\begin{rotate}{90}
$L(\stau_1)\times (\frac{m_{\stau_1}}{\rm 100 GeV})^2$ [$\mu m$]
\end{rotate}
\vskip32mm
\hskip105mm
$\Delta m^2_{\odot}$ [eV$^2$]
\vskip0mm
\caption{Stau decay length times $(\frac{m_{\stau_1}}{\rm 100 GeV})^2$ 
  in [$\mu m$] versus solar mass squared 
  difference $\Delta m_{\odot}^2$ [$eV^2$] in scenario I, defined as 
  $\lambda_{ijk}$ being responsible for the solar mass scale. If the
  trilinear loop is responsible for the solar mass, the stau will
  decay with a length too small to be measured.}
\label{fig:Sc1Len}
 \end{figure}

If the parameters in Eq. \eqref{eq:mlambda} are indeed such, that the
solar mass squared difference is correctly explained, the scalar tau
will decay with a very short decay length, as demonstrated in Fig.
\ref{fig:Sc1Len}.  Note that for a possible linear collider currently
one expects to be able to measure decay lengths down to 10 $\mu$m
\cite{Behnke:2001qq} in an event-by-event analysis. For typical
expected luminosities \cite{Brinkmann:2001qn} and production cross
sections for charged scalars as calculated in \cite{Hirsch:2002ys} the
expected number of events is about $10^4$ events per year, implying
that around 10 events decay with length $\sim$ 10 $\mu$m if its mean 
value is $\sim$ 1.5 $\mu$m. In view of the small decay lengths
implied by this scenario (Fig. \ref{fig:Sc1Len}) we think that attempts
by experimentalists at improving the measurement of very short decay
lengths would be very interesting.

For other $\lambda_{ijk}$ to provide the solar mass scale, the decay
lengths of the corresponding sleptons would have to be even shorter.
In this sense, scenario I is a ``worst-case'' scenario, since only
ratios of $\lambda_{ijk}$ - and a {\em lower limit} on the absolute
values of these parameters - could be measured. However, it is
possible to turn this argument around and state that, if a finite
decay length for $\se_1$ and $\smu_1$ is found $\lambda_{ij1}$ and
$\lambda_{ij2}$ {\em can not contribute significantly} to the neutrino
mass matrix.
\begin{figure}[t]
\begin{tabular}{cc}
\includegraphics[width=0.45\textwidth,height=0.35\textwidth]{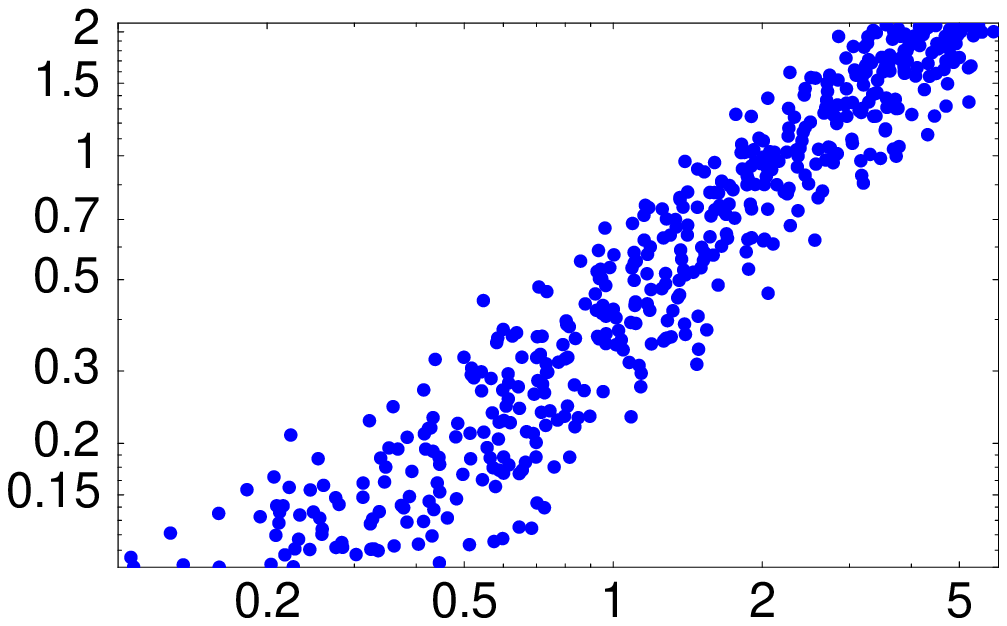}&
{\hskip5mm
\includegraphics[width=0.45\textwidth,
  height=0.35\textwidth]{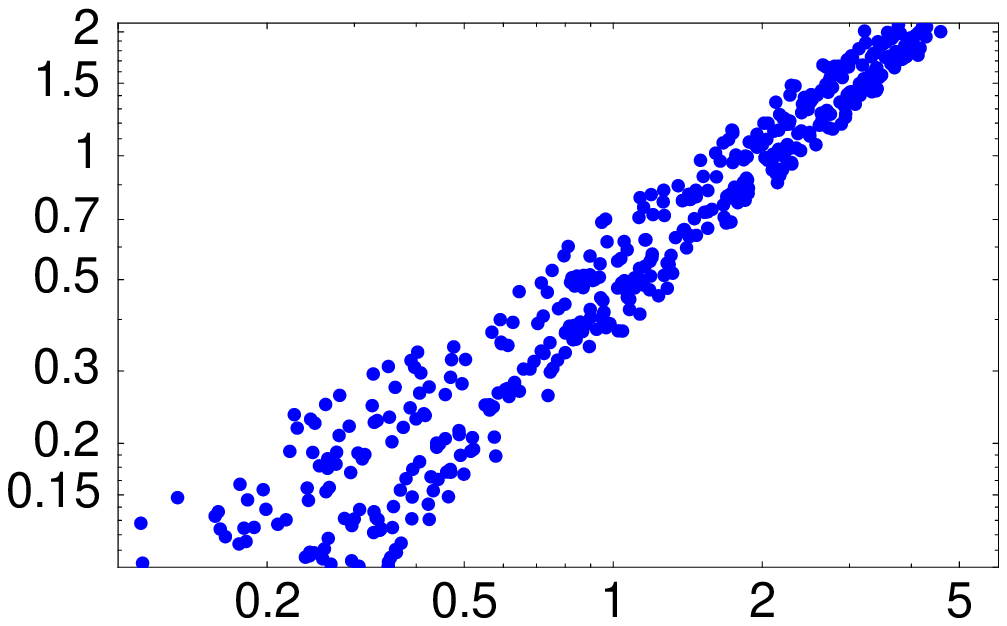}}
  \end{tabular}
\vskip-4mm
\hskip0mm
\begin{rotate}{90}
{\small $(1-2 \times Br^{\stau}_2)/(1-2 \times Br^{\stau}_1)$}
\end{rotate}
\vskip-4mm
\hskip70mm
\begin{rotate}{90}
{\small $(1-2 \times Br^{\stau}_2)/(1-2 \times Br^{\stau}_1)$}
\end{rotate}
\vskip3mm
\hskip50mm $\tan^2\theta_{\odot}$ \hskip55mm $\tan^2\theta_{\odot}$ 
\vskip0mm
\caption{Ratio of branching ratios 
  $(1-2 \times Br^{\stau}_2)/(1-2 \times Br^{\stau}_1)$ versus the
  solar angle, $\tan^2\theta_{\odot}$. The figure to the left is for
  $\forall \lambda_{ijk}$ non-zero and of similar magnitude, while the
  figure to the right assumes $\lambda_{123}=0$.}
\label{fig:Sc1Ang}
 \end{figure}

With a decay length of $\stau_1$ too small to be measured, only a consistency 
check of this scenario is possible. This check is provided by the measurement 
of a certain ratio of branching ratios, which is correlated with the solar 
angle, as demonstrated in Fig. (\ref{fig:Sc1Ang}). We want to stress, however, 
that this measurement can not distinguish trilinear terms from a bilinear-only 
world, since the latter leads to a very similar prediction for the branching 
ratios. Instead only disagreement between 
prediction and measurement would establish that neither scenario I nor 
bilinear loops are responsible for the solar neutrino mass scale.

\subsection{Scenario II: $\lambda_{ijk} \ll \lambda'_{ijk}$}

We now turn to the case where $\lambda_{ijk} \ll \lambda'_{ijk}$. If
the $\lambda'_{ijk}$ do not follow an {\em inverse} hierarchy, then
the main contribution of the trilinears to the neutrino mass,
Eq.~(\ref{eq:Triloopmass}), is approximately given by
\begin{eqnarray} \label{eq:mlambdap}
m^{\lambda'}\approx -\frac{3}
{16\pi^2} m_{b}
\sin 2\theta_{\ti b}
\ln\left(\frac{m^2_{{\ti b}_2}}{m^2_{{\ti b}_1}}\right)
\left(\hskip -2mm \begin{array}{ccc}
\lambda'^2_{133} 
\hskip -1pt&\hskip -1pt
\lambda'_{133} \lambda'_{233}
\hskip -1pt&\hskip -1pt
\lambda'_{133} \lambda'_{333}~ \\
\lambda'_{133}\lambda'_{233}
\hskip -1pt&\hskip -1pt
\lambda'^2_{233} 
\hskip -1pt&\hskip -1pt
\lambda'_{233}\lambda'_{333}~ \\
\lambda'_{133}\lambda'_{333} 
\hskip -1pt&\hskip -1pt 
\lambda'_{233}\lambda'_{333}
\hskip -1pt&\hskip -1pt
\lambda'^2_{333}
\end{array}\hskip -3mm \right)~,
\end{eqnarray}
since the $\ti b_i- b$ loop is largest due to the larger left-right
mixing expected in the sbottom sector and $m_d \ll m_s \ll m_b$.  Note
that $m^{\lambda'}$ has also a projective structure, similar to
$m_{\rm eff}$ in Eq.~\eqref{eq:meff}, but with different parameters.

\noindent
There are two limiting case, which we will study in some detail:

\begin{itemize}
\item Scenario IIA: Bilinear R-parity breaking responsible for 
solar physics, $\lambda'_{i33}$ for the atmospheric mass scale
\item Scenario IIB: Bilinear R-parity breaking responsible for 
atmospheric physics, $\lambda'_{i33}$ for the solar mass scale
\end{itemize}

\noindent
Consider scenario IIA first.  Here, due to the special structure of
Eq. \eqref{eq:mlambdap}, $\tan\theta_{23}$ and $\tan\theta_{13}$ can
be approximately expressed as
\begin{equation} \label{eq:thetaprth23}
\tan\theta_{23}\approx -
\frac{\lambda'_{233}}{\lambda'_{333}}~,
\end{equation}
\begin{equation} \label{eq:thetaprth13}
\tan\theta_{13}\approx -
\frac{\lambda'_{133}}{\sqrt{(\lambda'_{233})^2+(\lambda'_{333})^2}}~.
\end{equation}
The ratio in Eq.~\eqref{eq:thetaprth23} squared can be related to the
observable
\begin{equation} \label{eq:obser}
\frac{(\Gamma^{\ti\mu}_h)_{33}}{(\Gamma^{\ti\tau}_h)_{33}}\simeq
\left(\frac{\sin\theta_{\ti\mu}~\lambda'_{233}
}{\sin\theta_{\ti\tau}~\lambda'_{333}}\right)^2\approx
\frac{m^2_{\mu}}{m^2_{\tau}}~\tan^2\theta_{23},
\end{equation}
where $(\Gamma^{\ti\ell}_h)_{33}$ denotes the partial width of 
$\ti\ell_1$ decaying into the final state $\bar t b$.
Due to the tiny selectron mixing we expect the number of events 
$N(\ti e_1\to \bar t b)\ll N(\ti\mu_1\to \bar t b)$, such that 
the corresponding relation for Eq. \eqref{eq:thetaprth13} will be impossible 
to test. 
\begin{figure}[t]
\begin{tabular}{cc}
\includegraphics[width=0.45\textwidth]{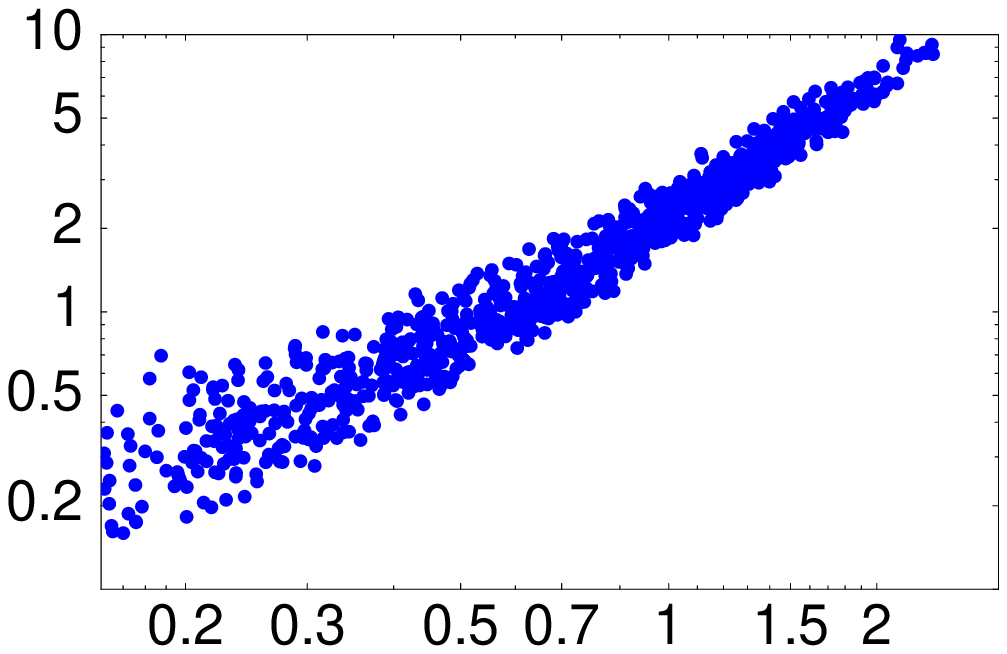} &
{\hskip5mm 
\includegraphics[width=0.45\textwidth]{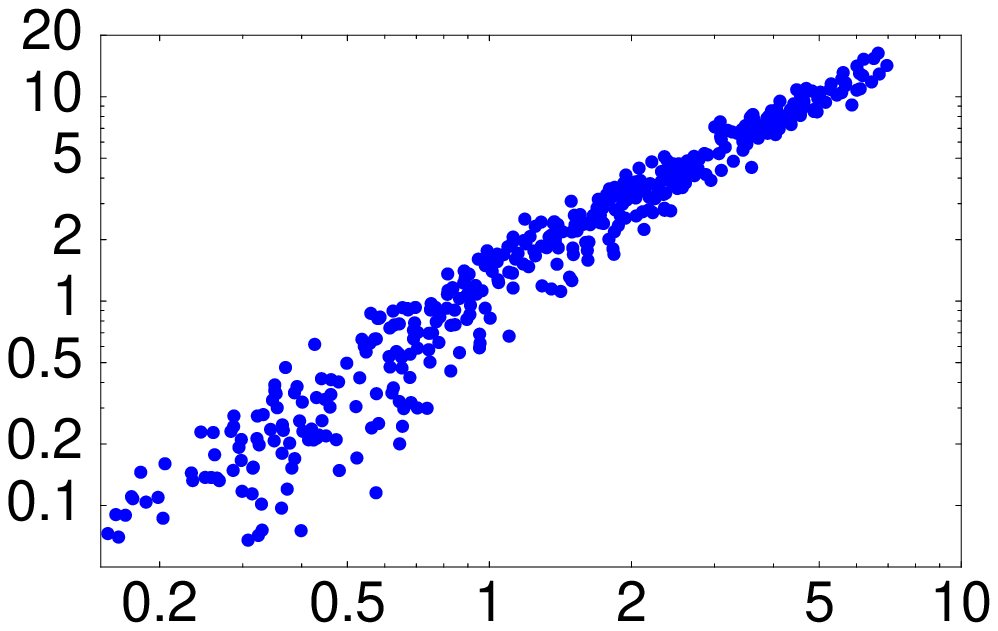}}
\end{tabular}
\vskip-25mm
\hskip0mm
\begin{rotate}{90}
$\tan^2\theta_{Atm}$
\end{rotate}
\vskip-5mm
\hskip70mm
\begin{rotate}{90}
$\tan^2\theta_{Atm}$
\end{rotate}
\vskip23mm
\hskip45mm $|\lambda'_{233}/\lambda'_{333}|$ 
\hskip35mm $(\Gamma^{\ti\mu}_h)_{33}/(\Gamma^{\ti\tau}_h)_{33}\times 10^3$
\vskip0mm
\caption{Ratio $\lambda'_{233}/\lambda'_{333}$ (left panel) and 
$(\Gamma^{\ti\mu}_h)_{33}/(\Gamma^{\ti\tau}_h)_{33}$ (right panel) versus 
the atmospheric angle, $\tan^2\theta_{Atm}$, for data points satisfying 
the criteria defined as scenario IIA. The plot to the right contains 
only points in which the decay ${\bar t}b$ is kinematically possible.}
\label{fig:Sc4Ang}
\end{figure}
\begin{figure}[t]
\begin{tabular}{cc}
\includegraphics[height=5cm,width=0.45\textwidth]{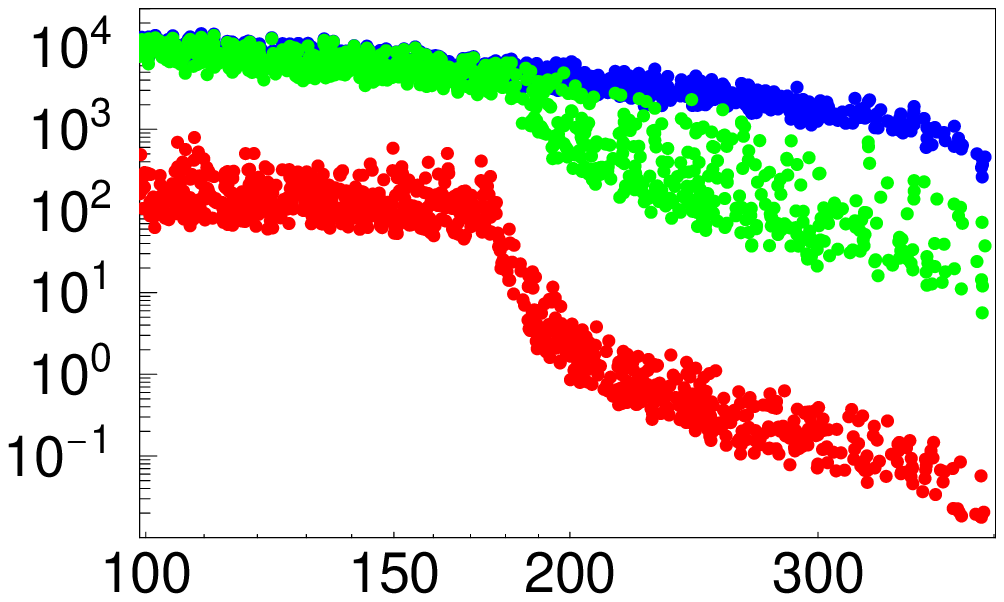} &
{\hskip5mm
\includegraphics[height=5cm,width=0.45\textwidth]{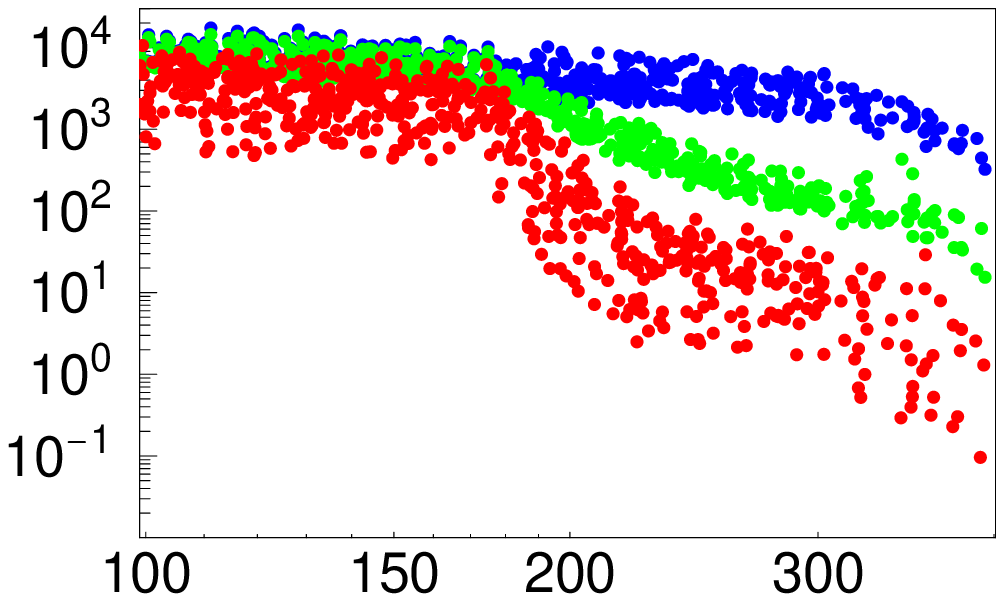}}
\end{tabular}
\vskip-7mm
\hskip0mm
\begin{rotate}{90}
$L(\stau_1)$, $L(\smu_1)$, $L(\se_1)$ [$\mu m$]
\end{rotate}
\vskip-5mm
\hskip70mm
\begin{rotate}{90}
$L(\stau_1)$, $L(\smu_1)$, $L(\se_1)$ [$\mu m$]
\end{rotate}
\vskip5mm
\hskip45mm $m_{\tilde l_1}$ [GeV]
\hskip50mm $m_{\tilde l_1}$ [GeV]
\vskip0mm
\caption{Decay lengths for $\stau_1$, $\smu_1$ and $\se_1$, 
  $L(\stau_1)$, $L(\smu_1)$, $L(\se_1)$ [$\mu m$], (bottom to top)
  versus the mass of the decaying particle, \hskip55mm $m_{\tilde
    l_1}$ [GeV]. On the left panel: Scenario IIA, right panel scenario IIB.
  The plots assume that only $\lambda'_{i33}$ are different from zero,
  see text.}
\label{fig:Sc4Len}
\end{figure}

For simplicity let us first consider the case in which only the 
$\lambda'_{i33}$ are non-zero. Modifications of the results for 
other $\lambda'_{ijk} \ne 0$ will be discussed at the end of this 
section.

In Fig. \ref{fig:Sc4Ang} we show the atmospheric angle versus 
$\lambda'_{233}/\lambda'_{333}$ (on the left panel) and 
$(\Gamma^{\ti\mu}_h)_{33}/(\Gamma^{\ti\tau}_h)_{33}$ (on the right panel) 
demonstrating to which accuracy one expects Eqs. \eqref{eq:thetaprth23} 
and \eqref{eq:obser} to work. Obviously Eq. \eqref{eq:obser} can be tested 
only if $m_{\stau_1}$ is bigger than $m_{t}+m_{b}$. 

Fig. \ref{fig:Sc4Len} (to the left) then shows the calculated decay
lengths for $\stau_1$, $\smu_1$ and $\se_1$ in scenario IIA. The top
threshold for the $\stau_1$ and $\smu_1$ decays is clearly visible,
while it is invisible for the selectron case due to the negligible
left-right mixing dictated by the small electron Yukawa coupling. 
This explains why the $\se_1$ decay length does not vary
much over the range of masses shown. This is due to the fact that the
decay of the $\se_1$ is completely dominated by the final state
$e\sum\nu$, induced by the bilinear term $|{\vec \Lambda}|$. For the
$\stau_1$ below threshold the decay is dominated by the final state
${\bar c}b$, which is induced from $\lambda'_{333}$ due to the
non-zero value of $V_{cb}$, with a branching ratio into $\tau\sum\nu$
being a few percent. Above threshold the branching into
${\bar t}b$ quickly grows to more than 99 \%, with the final state
${\bar c}b$ being suppressed by a factor of $V_{cb}^2$ and the final
state $\tau\sum\nu$ is tiny. For the decay of the $\smu_1$, on the
other hand, below threshold $\mu\sum\nu$ dominates, with a branching 
ratio into ${\bar c}b$ of about $\sim$ (1-20) \%, while above 
threshold ${\bar t}b$ 
becomes quickly dominant, but (some) percent of the decays still go to
$\mu\sum\nu$.  Decays $\tilde{e}_{i1} \to e_j \sum\nu$ for $i \ne j$ are
always very small, due to the assumed smallness of $\lambda_{ijk}$.

From Fig. \ref{fig:Sc4Len} one concludes that an experimental test of
Eq. \eqref{eq:obser} might be possible, but the decay lengths of the
$\stau_1$ are so short such that either (i) an improvement in
detecting short decay lengths will be necessary or (ii) additional
input from the sbottom sector is needed. \footnote{Fixing the sbottom
  mixing angle and masses would allow to {\em calculate} the absolute
  size of $\lambda'_{333}$ needed for scenario IIA, thus fixing the
  total $\stau_1$ width.}

Turning to scenario IIB, in the right panel of Fig. \ref{fig:Sc4Len}
we plot the calculated decay lengths of the charged scalar leptons as a
function of the charged scalar mass. The main difference to scenario
IIA is that the $\stau_1$ now decays with a visible length, apart from
some exceptional points where $m_{\stau_1}$ is very close to
$\sqrt{s}/2$. This can be traced to the fact that smaller values of
$\lambda'_{i33}$ are needed in this scenario since the solar mass
scale is smaller than the atmospheric one. Note that the plot assumes
that all $\lambda'_{i33}$ are of the same order of magnitude. While
$\lambda'_{333}$ can not be much larger than the values used in this
plot without violating the main assumption of scenario IIB, it could
well be smaller than $\lambda'_{133}$ or $\lambda'_{233}$, since
non-zero values for the latter are sufficient to create a non-zero
solar neutrino mass difference (and mixing). We note in passing that 
in the (unlikely)
case of $\lambda'_{333}=0$ the $\stau_1$ decay lengths would approach
the $\se_1$ decay lengths shown, with the dominating final state being
$\tau\sum\nu$ (again due to the assumed absence of $\lambda_{ij3}$).

Let us finally discuss modifications of the results discussed for
scenarios IIA and IIB once we allow for other $\lambda'_{ijk} \ne 0$.
Non-zero $\lambda'_{ijk}$ would, of course, affect the branching
ratios discussed above. Since final states ${\bar q}_jq_k$ can not be
distinguished experimentally for $j=1$ and $k=1,2$, only the sum of the
corresponding $(\lambda'_{ijk})^2$ are measurable. More important for
us, however, is that additional final states will shorten the decay
lengths of the sleptons compared to the ones shown in Fig.
\ref{fig:Sc4Len}.  Numerically we find that in scenario IIA the decay
length of the $\smu_1$ is always visible unless $\lambda'_{2jk}$ for
$j,k \ne 3$ are several times {\em larger} than the values for
$\lambda'_{233}$ needed to explain the atmospheric neutrino mass difference. 
The situation is different for $\stau_1$. Here, values of $\lambda'_{3jk}$
for $j,k \ne 3$ smaller than the ``correct'' $\lambda'_{333}$ by a
factor of a few are needed, otherwise the decay length of the
$\stau_1$ becomes to short to be realistically measurable. In scenario
IIB similar comments apply, although in this case even larger ratios 
$\lambda'_{ijk}/\lambda'_{233}$ are allowed for a visible decay 
length of the $\smu_1$ decays. For the
$\stau_1$ in scenario IIB $\lambda'_{3jk}$ as big as $\lambda'_{333}$
are possible for decay lengths to be observable. The decay of the
$\se_1$, on the other hand, never shows any sensitivity to
$\lambda'_{1jk}$, again due to the tiny left-right mixing in the
selectron sector.

\section{Conclusions}

We have studied the decay properties of a right charged scalar
lepton LSP (CSLSP) in a general model of R-parity violation containing
both bilinear and trilinear terms. Branching ratios and decay lengths
of CSLSPs contain information about ratios and absolute values of
R-parity violating couplings which at the same time contribute to the
neutrino mass matrix. We have investigated to what extent it is
possible to test experimentally whether bilinear or trilinear terms
give the dominant contribution to the neutrino masses.

Due to the huge number of parameters characterizing trilinear models
the resulting phenomenology can be quite diverse. In fact one can
envisage a mixed situation in which
bilinear terms are responsible for generating the neutrino masses
required to account for current oscillation data, yet the CSLSP decays
are governed by trilinears. Thus in general the complexity of the
physics resulting from Eq.~(\ref{eq:rpvpot}) is such that the
existence of non-zero bilinear terms can not be established.

Nevertheless, we have shown that different scenarios for neutrino
masses exist which lead to considerably different phenomenology and thus 
allows for the origin of neutrino mass to be experimentally probed. In
some cases we find that bilinears and trilinears can be clearly
distinguished.  Especially noteworthy is that in trilinear-only models
all right CSLSPs should obey $Br(\ti\ell_1\to (e,\mu,\tau) \sum
\nu_i) < 0.5$.  This is to be contrasted with the bilinear model
predicting \cite{Hirsch:2002ys} $Br(\se_1 \to e \sum \nu_i) \simeq 1$.

Decay lengths for the CSLSPs could be measurably large or very small
depending mainly on the absolute values of $\lambda_{ijk}$. Observing
a finite length for $\se_1$ and/or $\smu_1$ would establish that the
corresponding $\lambda_{ij1}$ and $\lambda_{ij2}$ do not contribute
significantly to the neutrino mass matrix. If $\lambda_{ijk}$ are
somewhat smaller than $\lambda'_{ijk}$ hadronic final states will have
measurable branching ratios, at least for the $\stau_1$ and depending
on $\lambda_{ij2}/\lambda'_{2jk}$ also for the $\smu$. In contrast,
note  that in the bilinear-only model \cite{Hirsch:2002ys} hadronic
final states are never visible.

Finally, since reasonable scenarios for the neutrino mass matrix exist
in which the decay length of the $\stau_1$ is just at or below the
borderline of what is currently thought of being experimentally
accessible \cite{Bartl:2003fh} we stress that efforts to optimize
decay length measurements might be worth undertaking.

\section*{Acknowledgments}

This work was supported by Spanish grant BFM2002-00345, by the
European Commission RTN grant HPRN-CT-2000-00148 and HPRN-CT-2000-00149, 
by the `Fonds zur F\"orderung der
wissenschaftlichen Forschung' of Austria FWF, Project No. P13139-PHY 
and No. P16592-N02 and by Acciones Integradas 
Hispano-Austriaca (HU2000-0019). M. H. is
supported by a Spanish MCyT Ramon y Cajal contract.  T.K. acknowledges
financial support from the European Commission Research Training Site
contract HPMT-2000-00124. W.~P.~is supported by the 'Erwin
Schr\"odinger fellowship No.~J2272' of the `Fonds zur F\"orderung der
wissenschaftlichen Forschung' of Austria FWF and partly by the Swiss
`Nationalfonds'.

\section{Appendix}
Here we give the solutions for the trilinear
couplings $\lambda^2_{ijk}$ by inverting Eq.~\eqref{eq:lep}
and making an expansion in $\theta_{\ti\ell}$ up to 
${\mathcal O}(\theta_{\ti\ell}^2)$.
This yields
\begin{eqnarray} \label{eq:tricoupe}
\lambda_{121}^2 = \frac{1}{2}~C^{-1}~\{
\Gamma^{\ti e}_{\rm tot}(1-2 Br^{\ti e}_3)+
\theta_{\ti e}^2(\Gamma^{\ti e}_{\rm tot}(Br^{\ti e}_2-Br^{\ti e}_3)
-\Gamma^{\ti\mu}_{\rm tot} Br^{\ti\mu}_1
+\Gamma^{\ti\tau}_{\rm tot} Br^{\ti\tau}_1)
\},
\nonumber\\[2mm]
\lambda_{131}^2 =
\frac{1}{2}~C^{-1}~\{
\Gamma^{\ti e}_{\rm tot}(1-2 Br^{\ti e}_2)+
\theta_{\ti e}^2(\Gamma^{\ti e}_{\rm tot}(Br^{\ti e}_3-Br^{\ti e}_2)
+\Gamma^{\ti\mu}_{\rm tot} Br^{\ti\mu}_1
-\Gamma^{\ti\tau}_{\rm tot} Br^{\ti\tau}_1)
\},
\nonumber\\[2mm]
\lambda_{231}^2 = 
\frac{1}{2}~C^{-1}~\{
\Gamma^{\ti e}_{\rm tot}(1-2 Br^{\ti e}_1)+
\theta_{\ti e}^2(\Gamma^{\ti e}_{\rm tot}(Br^{\ti e}_3+Br^{\ti e}_2)
-\Gamma^{\ti\mu}_{\rm tot} Br^{\ti\mu}_1
-\Gamma^{\ti\tau}_{\rm tot} Br^{\ti\tau}_1)
\},
\end{eqnarray}
\begin{eqnarray} \label{eq:tricoupm}
\lambda_{122}^2 = 
\frac{1}{2}~C^{-1}~\{
\Gamma^{\ti\mu}_{\rm tot}(1-2 Br^{\ti\mu}_3)+
\theta_{\ti\mu}^2(\Gamma^{\ti\mu}_{\rm tot}(Br^{\ti\mu}_2-Br^{\ti\mu}_3)
-\Gamma^{\ti e}_{\rm tot} Br^{\ti e}_2
+\Gamma^{\ti\tau}_{\rm tot} Br^{\ti\tau}_2)
\},
\nonumber\\[2mm]
\lambda_{132}^2 = 
\frac{1}{2}~C^{-1}~\{
\Gamma^{\ti\mu}_{\rm tot}(1-2 Br^{\ti\mu}_2)+
\theta_{\ti\mu}^2(\Gamma^{\ti\mu}_{\rm tot}(Br^{\ti\mu}_2+Br^{\ti\mu}_3)
-\Gamma^{\ti e}_{\rm tot} Br^{\ti e}_2
-\Gamma^{\ti\tau}_{\rm tot} Br^{\ti\tau}_2)
\},
\nonumber\\[2mm]
\lambda_{232}^2 = 
\frac{1}{2}~C^{-1}~\{
\Gamma^{\ti\mu}_{\rm tot}(1-2 Br^{\ti\mu}_1)+
\theta_{\ti\mu}^2(\Gamma^{\ti\mu}_{\rm tot}(Br^{\ti\mu}_3-Br^{\ti\mu}_2)
+\Gamma^{\ti e}_{\rm tot} Br^{\ti e}_2
-\Gamma^{\ti\tau}_{\rm tot} Br^{\ti\tau}_2)
\},
\end{eqnarray}
\begin{eqnarray} \label{eq:tricoupt}
\lambda_{123}^2 = 
\frac{1}{2}~C^{-1}~\{
\Gamma^{\ti\tau}_{\rm tot}(1-2 Br^{\ti\tau}_3)+
\theta_{\ti\tau}^2(\Gamma^{\ti\tau}_{\rm tot}(Br^{\ti\tau}_1+Br^{\ti\tau}_2)
-\Gamma^{\ti e}_{\rm tot} Br^{\ti e}_3
-\Gamma^{\ti\mu}_{\rm tot} Br^{\ti\mu}_3)
\},
\nonumber\\[2mm]
\lambda_{133}^2 = 
\frac{1}{2}~C^{-1}~\{
\Gamma^{\ti\tau}_{\rm tot}(1-2 Br^{\ti\tau}_2)+
\theta_{\ti\tau}^2(\Gamma^{\ti\tau}_{\rm tot}(Br^{\ti\tau}_1-Br^{\ti\tau}_2)
-\Gamma^{\ti e}_{\rm tot} Br^{\ti e}_3
+\Gamma^{\ti\mu}_{\rm tot} Br^{\ti\mu}_3)
\},
\nonumber\\[2mm]
\lambda_{233}^2 = 
\frac{1}{2}~C^{-1}~\{
\Gamma^{\ti\tau}_{\rm tot}(1-2 Br^{\ti\tau}_1)+
\theta_{\ti\tau}^2(\Gamma^{\ti\tau}_{\rm tot}(Br^{\ti\tau}_2-Br^{\ti\tau}_1)
+\Gamma^{\ti e}_{\rm tot} Br^{\ti e}_3
-\Gamma^{\ti\mu}_{\rm tot} Br^{\ti\mu}_3)
\},
\end{eqnarray}
where $\Gamma^{\ti\ell}_{\rm tot}$ denotes the total leptonic decay
width of the appropriate scalar lepton, which is, for the case where
$\ti\ell_1\simeq \ti\ell_R$, not much different from the total decay
width if $\lambda'\lsim \lambda$.  $C \equiv \frac{m_{\ti\ell_1}}{16\pi}$, 
where we have made use of the fact that
$m_{\ti e_1}\simeq m_{\ti\mu_1} \simeq m_{\ti\tau_1}(\equiv
m_{\ti\ell_1})$.  As can be seen in
Eqs.~\eqref{eq:tricoupe}-\eqref{eq:tricoupt} the next to leading term
is only of the order $\theta_{\ti\ell}^2$.  The measurement of
$\Gamma^{\ti\ell}_{\rm tot}$ and $\theta_{\ti\ell}$ provides the
information whether the expansion to ${\mathcal
  O}(\theta_{\ti\ell}^2)$ is sufficient.

\end{document}